\documentclass[review]{elsarticle}

\usepackage{lineno,hyperref}
\usepackage{amsmath}
\usepackage{amsfonts}
\usepackage{amssymb}
\usepackage{graphicx}
\usepackage{float}
\usepackage{mwe}
\usepackage{subfig}
\usepackage{multirow}
\usepackage{scalefnt}
\usepackage{verbatim}
\usepackage[T1]{fontenc}
\usepackage{setspace}
\usepackage{verbatim}
\begin{document}

\begin{frontmatter}

\title{How Does ENSO Impact the Solar Radiation Forecast in South America? The Self-affinity Analysis Approach}







\author[label1]{T.B. Murari \corref{mycorrespondingauthor}} \cortext[mycorrespondingauthor]{Corresponding author}\ead{mura.learning@gmail.com}
\author[label1]
{A.S. Nascimento Filho}\ead{aloisio.nascimento@gmail.com}
\author[label1,label2]{M.A. Moret}\ead{mamoret@gmail.com}
\author[label1]{S. Pitombo}\ead{sergio.pitombo@fieb.org.br}
\author[label1]{A.A.B. Santos}\ead{alex.santos@fieb.org.br}
\address[label1]{Centro Universit\'ario Senai Cimatec, Salvador, BA, Brazil}
\address[label2]{Universidade do Estado da Bahia, Salvador, BA, Brazil}

\begin{abstract}

The major challenge we face today in the energy sector is to meet the growing demand for electricity with less impact on the environment. South America is an important player in the renewable energy resource. Brazil accelerated the growth of photovoltaic installed capacity in 2018. From April  of 2017 to April of 2018, the capacity increased $1351.5\%$. It is expected to reach the value of $2.4\ GW$ until the end of the year. The new Chilean regulation request that 20\% of the total electricity production in 2025 must come from renewable energy sources. The aim of this paper is to establish time series behavior changes between El Ni\~no Southern Oscillation and the solar radiation resource in South America. The results can be used to validate forecasts of energy production for new solar plants. The method used to verify the behavior of the time series was the Detrended Fluctuation Analysis. Solar radiation data were collected in twenty-five cities distributed inside the Brazilian solar belt, plus six cities in Chile, covering the continent from east to west, in a region with high potential of solar photovoltaic generation. The results shows the impact of El Ni\~no Southern Oscillation on the climatic behavior of the evaluated data. It is a factor that may lead to the wrong forecast of the long term potential solar power generation for the region.

\end{abstract}


\end{frontmatter} 


\section{Introduction}

Renewable energy sources can play an important role in solving the dilemma of increasing energy production capacity by minimizing interference in the environment ~\cite{millais2005}. Some studies have been investigating the potential contribution of renewable energy supplies to the global grid. It indicates that, in the second half of this century, their contribution might range from 20\% to more than 50\%, once we established the correct government policies \cite{ezzati2004}.\\

Regarding renewable energy investments, Brazil and Chile stand out in South America continent. Both are the first and second destination countries for foreign investment, in clean energy asset finance, in the region, attracting the most investment from overseas financiers from 2010 up to 2016. Brazil received US\$ $12.78$ billions and Chile, US\$ $6.92$ billions. For reference, in the same period, Latin America received a total of US\$ $36.1$ billions in clean energy investment from overseas funders \cite{climatescope2017}.\\

The installed capacity of power generation in Brazil reached $150.4$ GW in 2016, an increase of $9.5$ GW in relation to 2015. Among the sources that stand out most are hydropower, with 64.5\%, and biomass, which held $9.3$\%. Considering this import source, the total power supply reached $156.3$ GW in 2016. Thus, $80.6$\% of Brazil's total installed energy capacity is renewable, while, in the world, the indicator is only $33$\%, close to $6450$ GW \cite{EPE2016}.\\

According to the monthly monitoring bulletin of the Brazilian Electric System, the solar photovoltaic source in Brazil grown rapidly in recent years. Brazil reach the end of 2018 with $2400$ MW. Further, the 10-year Energy Expansion Plan (PDE 2024) estimates that the installed capacity of solar generation in Brazil will reach $8300$ MW by 2024. For instance, the Brazilian installed solar energy capacity increased $1351.5$\% from April of 2017 to 2018 \cite{MMEa}.\\

In Brazil, a Certified Energy Production Estimative (CEPE) for the future solar plant must be submitted to the government, by the interested company who wants to participate in auctions to generate this photovoltaic energy. This estimation is calculated based on, at least, twelve consecutive months of measurements within a radius of up to $10km$ from the project. In addition to horizontal solar radiation, data on temperature, relative air humidity and wind speed need to be collected. The rate of data loss must be less than $10$\% of the total and the absence of continuous measurements can not exceed fifteen days \cite{EPE2016}. According to \cite{palz2013}, solar radiation and temperature are the two main factors that influence energy production by photovoltaic modules.\\

The local measured data is compared with the same period data obtained from satellite models to obtain the Typical Meteorological Year, that is used in the calculation of CEPE. Moreover, these certified data will be correlated with a dataset collected from the closest weather station, for 10 years or more \cite{EPE2016}.\\

\begin{figure}[H]
\centering
\includegraphics[width=12cm]{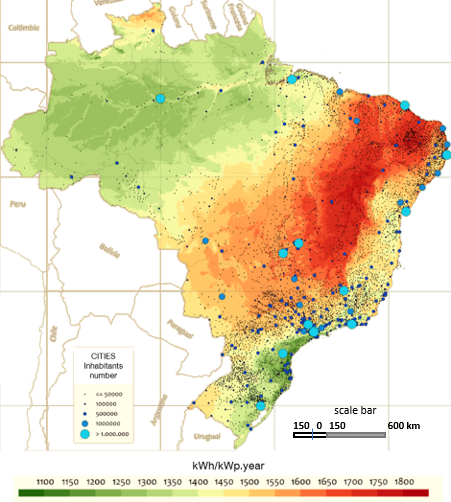}
\caption{\label{fig:solarmap} Photovoltaic power generation potential in Brazil. The size of the blue circles represent the number of inhabitants in each Brazilian city. Adapted from \cite{inpeinstituto2017}}
\end{figure}

Brazil has great potential for photovoltaic power generation. The Northeastern region has the highest potential of the country, with a mean value of the solar Global Horizontal Irradiance (GHI) of $5.9$ $kWh/m^{2}$. This high rate in the Northeast is explained by the low cloudiness condition \cite{inpeinstituto2017}. The Figure \ref{fig:solarmap} shows the map for photovoltaic power generation. It presents the maximum annual energy yield (measured in $kWh$ of annual energy generated by $kWp$ of installed photovoltaic energy) for the entire Brazilian territory.\\

Northern Chile (Atacama Desert) has a GUI of $3300$ $kWh/m^2$ on latitude tilt surfaces \cite{fthenakis2014}. In 2014, a new solar plant project in Atacama region start, and it will be the largest solar project in Latin America, to be used particularly for mining companies with operations in the region \cite{ITD2016}. The country can meet the $20$\% target of its capacity from renewable sources, before the 2025 deadline \cite{CER2016}, even in the absence of government subsidies \cite{ITD2016}. Beside that, Chile was the first country in Latin America to create a carbon tax. The Congress passed the called green tax in September of 2014 \cite{ITD2016}.\\

South America is impacted by the El Ni\~no Southern Oscillation (ENSO). Both El Ni\~no and La Ni\~na climate phenomenas are known as ENSO and they are the opposite phases of a natural climate pattern throughout the tropical Pacific Ocean ecosystem, which oscillates every 3 to 7 years \cite{noa2018,Kevin1997}. These events lead to significant differences in the average temperature of the oceans, winds, surface pressure and precipitation in parts of the Tropical Pacific Ocean \cite{noa2018}.\\

Brazil has an energy matrix with a renewable-thermal configuration, but, essentially, depends on a large hydropower plants, which can be an issue in times of severe drought. For instance, Brazil had some severe drought on Amazon region in the years of 2005 \cite{Phillips2009}, 2010 \cite{Lewis2011} and 2016 \cite{erfanian2017}. The same drought condition was experienced by Northeastern region in 2005, 2007, 2010, 2012 and 2016 \cite{erfanian2017,marengo2017}. It is one factor that may justify the current growth of non-hydropower plants in the country \cite{MMEa}.\\

El Ni\~no is associated with above average rainfall in central Chile during winter and late spring. The La Ni\~na is related to the below average rainfall in the same period and region. El Ni\~no is dry and La Ni\~na wet in southern-central Chile during the summer \cite{montecinos2003}. There is also a study about the techno-economical impact in the wind resource of Chile \cite{watts2017}.\\

Table \ref{Tab:ocorrencia} shows the occurrences of El Ni\~no and La Ni\~na in the world and their intensities based on the Oceanic Ni\~no Index values from 1950 to 2018. Events were classified as weak, moderate, strong or very strong \cite{cai2015,wang2002}.\\

\begin{table}[H]
\center
\caption{\label{Tab:ocorrencia} El Ni\~no and La Ni\~na occurrences \cite{noa2018}}
\begin{tabular}{cccc|ccc}
\hline
\multicolumn{4}{c|}{\textbf{El Ni\~no}} & \multicolumn{3}{c}{\textbf{La Ni\~na}} \\
\textbf{Weak} & \textbf{Moderate} & \textbf{Strong} & \textbf{Very Strong} & \textbf{Weak} & \textbf{Moderate} & \textbf{Strong} \\
\hline
1952-53 & 1951-52 & 1957-58 & 1982-83 & 1954-55 & 1955-56 & 1973-74 \\
1953-54 & 1963-64 & 1965-66 & 1997-98 & 1964-65 & 1970-71 & 1975-76 \\
1958-59 & 1968-69 & 1972-73 & 2015-16 & 1971-72 & 1995-96 & 1988-89 \\
1969-70 & 1986-87 & 1987-88 &  & 1974-75 & 2011-12 & 1998-99 \\
1976-77 & 1994-95 & 1991-92 &  & 1983-84 &  & 1999-00 \\
1977-78 & 2002-03 &  &  & 1984-85 &  & 2007-08 \\
1979-80 & 2009-10 &  &  & 2000-01 &  & 2010-11 \\
2004-05 &  &  &  & 2005-06 &  &  \\
2006-07 &  &  &  & 2008-09 &  &  \\
2014-15 &  &  &  & 2016-17 &  & \\
\hline
\end{tabular}
\end{table}

South America is a drought hot spot in some future weather projections because of its potential to drastically react to excessive warming and drying \cite{marengo2017,cox2008}, and El Ni\~no events are important predictors for severe droughts over the Brazilian Amazon and Northeast \cite{hastenrath1984,marengo2004}. The greatest analyzed drought, between 1982 and 2017, is the known 2016 drought, during the El Ni\~no, an unprecedented dry period \cite{erfanian2017}.\\

Significant local studies on the effects of the ENSO in various parts of the globe have been important in establishing the prediction of solar energy \cite{Davy,mohammadi2018,chang2017}. \cite{Dasilva} provides an overview study of the decrease of solar radiation for four climatic zones of northeastern Brazil, which can be attributed to the global dimming effect influenced by ENSO. On the other hand, the variability of the solar irradiation in the Atacama desert is influenced by the ENSO. These phenomena will result in years with significantly different solar irradiation than TMY \cite{Bravol}. The time series solar radiation exhibit characteristics associated with natural climatic phenomena with example ENSO, thus revealing that the Descending Solar Radiation can be considered as an indirect indicator for these \cite{Papadimas}.\\

It brings out the following question: Does ENSO uniformly impact the solar radiation and, consequently, the forecast of solar power production in South America? The aim of this paper is to establish time series behavior changes between ENSO and the solar radiation resource spatially. These relationships may be used to guide forecasts for new solar energy plants. Therefore, the scope of this work is limited to the evaluation of the solar radiation variable.\\

\section{Materials and methods}

\subsection {Dataset Acquisition}

\begin{figure}[H]
	\begin{center}
		\includegraphics[scale=1.15]{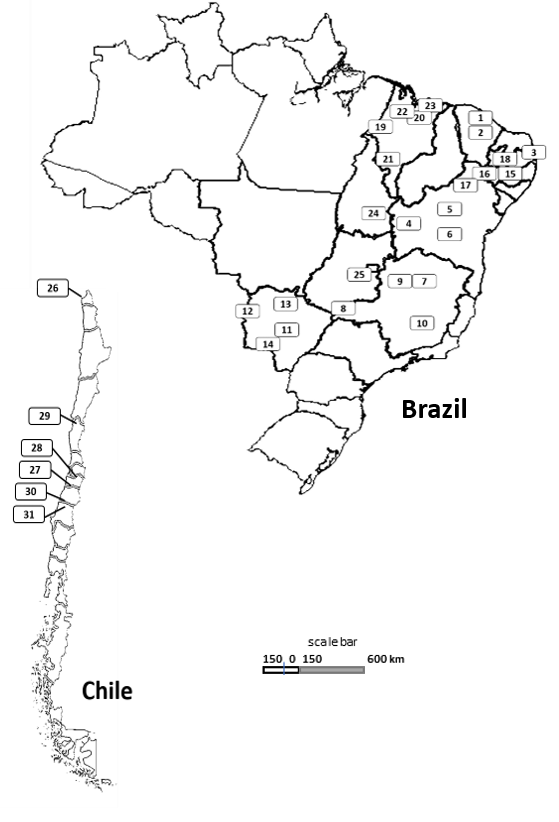}
	\end{center}
	\caption{The accumulated solar radiation were collected from thirty one in Brazil and Chile within the high potential region for solar energy production.}
	\label{figura:Cities_solar_map}
\end{figure}

\begin{figure}[H]
	\begin{center}
		\includegraphics[scale=0.5]{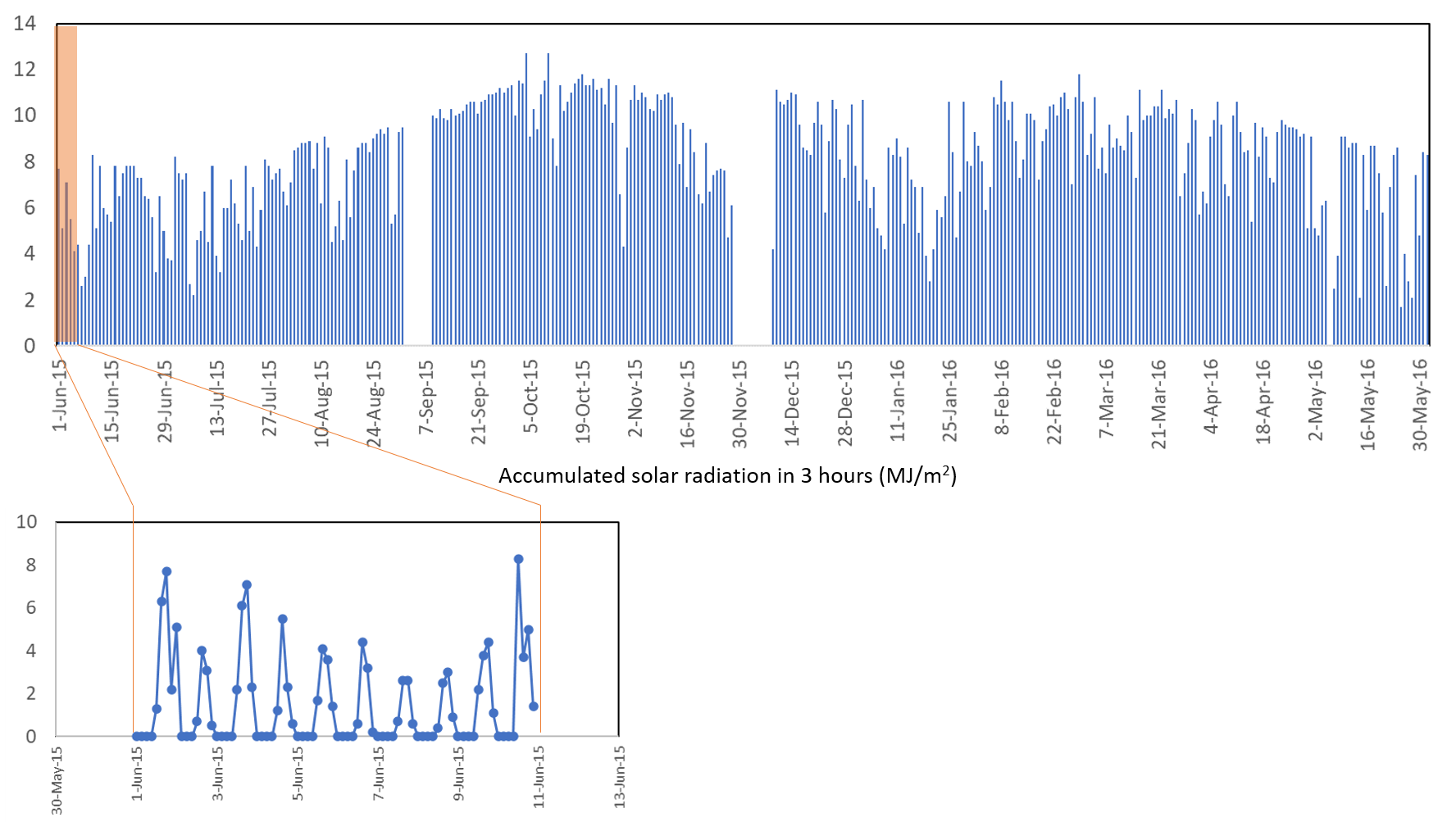}
	\end{center}
	\caption{Raw time series from Piat\~a, Bahia, during the El Ni\~no period}
	\label{figura:time_series_raw}
\end{figure}

All measurements represents the accumulated solar radiation per $3$ hours ($MJ/m^2$) and it were provided by the Brazilian Center for Weather Forecasting and Climate Studies (CPTEC) of the National Institute for Space Research (INPE) \cite{SINDA} and the National Agro-climatic Network (Agromet) of Chile \cite{INIA}. This data was automated collected from weather stations that are widely distributed in several regions of Brazil, within the solar belt \cite{pereira2017}, and Chile. The chosen region has a huge potential for solar power generation. It was evaluated datasets from thirty one cities that reach a maximum threshold of 10\% of lost data inside this region (Figure \ref{figura:Cities_solar_map} and Table \ref{table:cities_states}). The lost data were removed from the city dataset. An example of these time series can be seen on \ref{figura:time_series_raw}, that represent the collected data from Piat\~a	during the El Ni\~no, between 2015 and 2016.\\

This threshold was defined based on the results behavior, that present a crossover from uncorrelated signal on short range to correlated signal on long range. According to \cite{ma2010}, the global scaling exponent of positively correlated signals ($1.5 \geq \alpha > 0.5$), remains unaffected even for data loss of up to 90\%, and shows no observable changes in the local scaling for up to 65\% of data loss. \cite{chen2002} found that remove a small segment of the data strongly affects anti-correlated signals, leading to a crossover from an anti-correlated regime at small scale to an uncorrelated regime at large scale.\\ 

These data were collected in three different periods of time, each one corresponding to a full year (2920 points maximum for year, disregarding lost data) as follow. The selected period coincides with a very strong occurrence of El Ni\~no and a strong occurrence of La Ni\~na \cite{noa2018}:

\begin{itemize}
\item La Ni\~na - from Jun 01, 2010 to Jun 01, 2011
\item Neutral - from Jun 01, 2013 to Jun 01, 2014
\item El Ni\~no - from Jun 01, 2015 to Jun 01, 2016
\end{itemize}

\begin{table}[H]
	\center
\caption{Cities of the weather stations where the analyzed data were collected. Tagged cities from 1 to 25 are inside Brazil and, from 26 to 31, inside Chile}
	\begin{tabular}{lll}
		\hline
        \textbf{Tag}	&	\textbf{State or Region}	&	\textbf{City}\\
		\hline 
        1	&	Cear\'a				&	Canind\'e			\\	
        2	&	Cear\'a				&	Quixeramobim		\\	
        3	&	Para\'iba			&	Capim			\\	
        4	&	Bahia				&	S\~ao Desid\'erio		\\	
        5	&	Bahia				&	Irec\^e				\\
        6	&	Bahia				&	Piat\~a				\\	
        7	&	Minas Gerais		&	Montes Claros		\\	
        8	&	Minas Gerais		&	Santa Vit\'oria		\\	
        9	&	Minas Gerais		&	Santa F\'e			\\
        10	&	Minas Gerais		&	Belo Horizonte		\\
        11	&	Mato Grosso do Sul	&	Campo Grande	\\
        12	&	Mato Grosso do Sul	&	Corumb\'a	\\
        13	&	Mato Grosso do Sul	&	Coxim \\
        14  &	Mato Grosso do Sul	&	Jardim	\\
        15	&	Pernambuco			&	Arcoverde\\
        16	&	Pernambuco			&	Bel\'em do S\~ao Francisco	\\
        17	&	Pernambuco			&	Petrolina	\\
        18	&	Pernambuco			&	S\~ao Jos\'e do Egito	\\
        19	&	Maranh\~ao			&	A{\c c}ail\^andia	\\
        20	&	Maranh\~ao			&	Coroat\'a	\\
        21	&	Maranh\~ao			&	Riach\~ao	\\
        22	&	Maranh\~ao			&	Santa In\^es	\\
        23	&	Maranh\~ao			&	Urbano Santos	\\
        24	&	Tocantins			&	Chapada da Natividade	\\
        25	&	Goi\'as				&	An\'apolis	\\
        26	&	Arica and Parinacota 	&	Lluta Bajo, Arica \\
        27	&	Maule				& Botalcura, Pencahue	\\
        28	&	O$'$Higgins			& 	El Tambo \\
        29	&	Coquimbo			&	Las Rojas, La Serena \\
        30	&	Maule				&	Los Despachos, Cauquenes \\
        31	&	B\'io B\'io			&	Coronel de Maule \\
		\hline
\label{table:cities_states}
\end{tabular}
\end{table}

\subsection{Self-affinity Analysis method}

The method called Detrended Fluctuation Analysis (DFA) \cite{Peng1994} was proposed  to analyze long-range
power-law correlations in nonstationary systems, and extended to higher order polynomials by \cite{kantelhardt2001,hu2001}. It have being widely applied in non-stationary time series including the following: transport \cite{filho2008}, combustion \cite{souza2015}, proteins \cite{figueiredo2010}, dengue fever \cite{azevedo2016}, astrophysical systems \cite{Moret2003}, sunspots \cite{Moret2014}, cloud structures evaluation \cite{Ivanova1999,Ivanova2000} and weather analyzes \cite{Anjos2015,Santos2012,govindan2004,kavasseri2005,koccak2009,kurnaz2004,chen2007,matsoukas2000,talkner2000}, including solar radiation \cite{Anjos2015,zeng2013,madanchi2017}.\\

DFA is calculated according to the following steps. The original time series $s_i$ , where  $s_i$ is the accumulated solar radiation ($MJ/m^2$) each $3$ hours, with $i=1,...,N$, and $N$ is the total number of measurements registered. The time series $s_i$ is integrated, where $\left \langle s \right \rangle $ is the average value of $s_i$.

\begin{equation}
y(k)=\sum_{i=1}^{k}[s_{i}-\left \langle s \right \rangle]  
\label{Yn}
\end{equation}

The integrated signal $y(k)$ is divided into non-overlapping boxes of equal length $n$; and $y(k)$ is fitted using a polynomial function, which represents the trend in this box. Then, $y(k)$ is detrended by subtracting the local trend $y_n(k)$ within each box. For a given size box, the root-mean-square fluctuation, $F(n)$, is calculated as 

\begin{equation}
F(n)=\sqrt{\frac{1}{N}\sum_{k=1}^{N}[y(k)-y_n(k)]^2}.
\label{Fn}
\end{equation}
\\

The equation \eqref{Fn} is repeated for a wide range of scales to estimate the relationship between $F(n)$ and the box size. The scaling exponent $\alpha $ is characterized by power law $F(n) \sim n^\alpha$. It means a self-affinity parameter expressing the long-range power-law correlation properties.\\

Furthermore, the scaling exponent $\alpha$ will be used to assess the long-range correlation influences on the future behavior. The $\alpha$ exponent is classified according to the following rules, as previously applied by \cite{peng1995,Malamud1999,galhardo2009,souza2015,nascimento2017,nascimento2018}: 

\begin{itemize}
\item anti-persistent signal ($0 < \alpha < 0.5$)
\item white noise with no memory ($\alpha = 0.5$)
\item persistent signal ($0.5 < \alpha < 1$)
\item noise type $1/f$ ($\alpha = 1$)
\item sub-diffusive process ($1 < \alpha$)
\item brown noise ($\alpha = 1.5$)
\end{itemize}

A positive correlation, in time series, means that an increasing trend in the past may be followed by an increasing trend in the future. It has a persistent signal. A negative correlation means that an increasing trend in the past may be followed by a decreasing trend in the future. It is called as anti-persistent \cite{delignieres2011}.\\

The normal diffusion mean squared displacement of diffusing particles has linear time dependence, one characteristic of Brownian motion and a result of the central limit theorem \cite{reveillac2009}, where all steps in the diffusion process have equal length and travel time. But a lot of dynamic systems mean squared displacement presents a non-linear growth over time. A long step may be very fast, called as super diffusion, or a short step may be very slow, known as sub-diffusion \cite{sharifi2012}. This anomalous diffusion, described by a power law, has a non-linear relation to time \cite{havlin2002}.

\section{Results and Discussion}

\begin{figure}[H]
	\begin{center}
		\includegraphics[scale=0.35]{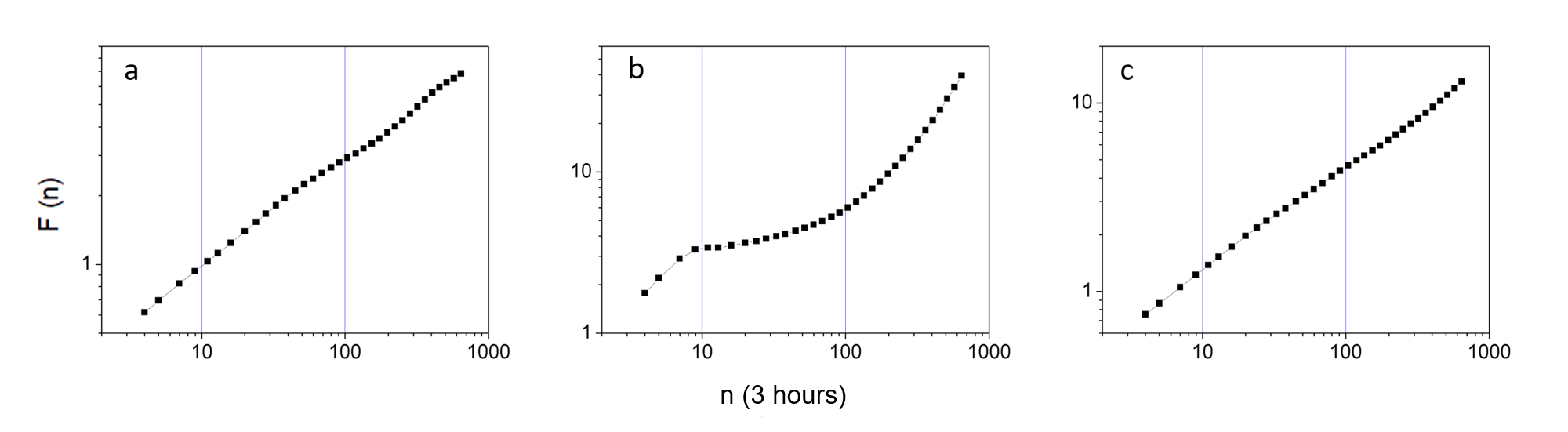}
	\end{center}
	\caption{Coxim (MS) is the city that presented the biggest range variation on the $\alpha$ for solar radiation, moving from a anti-persistent state on the (a) El Ni\~no to a sub-diffusive one during the (b) La Ni\~na, and become persistent in the (c) neutral year, for a time scale window greater than ten days.}
	\label{fig:COXIM_DFA_GRAPH}
\end{figure}

The results of the self-affinity evaluation indicated, through the correlation exponent $\alpha$, the presence of crossover, as seen in figure \ref{fig:COXIM_DFA_GRAPH} (b). Crossover is a change point in a scaling law, where one scaling exponent applies for small scale parameters and another scaling exponent applies for large scale parameters \cite{peng1995,kantelhardt2009}.\\

We used the second derivative of the F(n) curve to accurately determine the crossover points. The first average crossover point for all curves is on $n = 9$. The second average crossover point for all curves lies between $n = 80$ (10 days) and $n = 104$ (13 days). According to \cite{lovejoy2013,lovejoy2015,lovejoy2019}, the transition from weather to macroweather is 10 days. Macroweather is dynamic regime in which fluctuations in atmospheric variables, like temperature and precipitation, reduce with timescale.\\

Based on derivative of the F(n) curve and the macrowether inner scale of 10 days, it was defined the initial and final scale to fit the F(n) curve, within each scale range, as follow:

\begin{itemize}
\item short range - $n \leq 9$ $(\leq 27$ hours)
\item weather - $9 < n\leq 80$ ($\leq 10$ days)
\item macroweather - $n > 80$ $(> 10$ days)
\end{itemize}

The presence of crossover were found in other studies. \cite{santos2019analysis} found at least two different scale exponents (using DFA) over the analyzed period, a subdiffusive process in small time scales and persistent in long time scales. It shows the presence of crossover in wind speed on the region of Abrolhos, Brazil. Also, \cite{santos2019analysis} says that wind energy is directly related to solar radiation, because the winds are generated by the non-uniform heating of the planet's surface. \cite{monetti2003long} studied the long-term persistence in the Atlantic and Pacific sea surface temperature fluctuations, for the period 1856 – 2001. They found that, in contrast to land stations, there exist two pronounced scaling regimes. In the short-time regime that roughly ends at 10 months, $\alpha$ in the northern Atlantic ($\approx$ 1.4) that differs from the other oceans ($\approx$ 1.2), This behavior is distinct from the temperature fluctuations on land, where alpha is close to 0.65, above typically 10 days. In Fernando de Noronha, Brazil, the wind speed and solar radiation dynamics presented persistent properties in long-term scale (alpha > 0.5) with stronger persistency for wind speed indicated by the higher value of scaling exponent \cite{dos2015long}.\\

Firstly, it was calculated $\alpha$ for short range, representing a box of twenty seven hours. For both Brazil and Chile, the coefficient is persistent in any period (ENSO or neutral), except for Capim ($0.49\pm0.01$) in the neutral year and Jardim ($0.49\pm0.01$) in the La Ni\~na year, both anti-persistent but very close to the persistent interval.\\

The second period is related to the DFA coefficient of solar radiation between twenty seven hours and ten days. We ran hypothesis tests to assess whether these data could be described as power laws. The biggest found Prob>F is $1.05623\cdot10^{-6}$ for the Neutral period of Petrolina and any Pearson coefficient is bigger than $0.9$. Chilean evaluated cities always presented a anti-persistent coefficient. In Brazil, the $\alpha$ for each city were essentially anti-persistent, but Capim ($0.57\pm0.02$) and Coxim ($0.54\pm0.01$) in the neutral season, Santa Vit\'oria ($0.55\pm0.03$) and Urbano Santos ($0.57\pm0.01$) in the El Ni\~no, and Jardim for any of the studied periods. \\

Further, the third calculated $\alpha$ represents a box of more than ten days, representing the macroweather scale. The neutral year were consistently persistent in Brazil, except for Campo Grande ($0.47\pm0.02$). Some sub-diffusive values were found for both evaluated ENSO years in S\~ao Desid\'erio, Santa Vit\'oria and Coxim.\\

\begin{figure}[H]
	\begin{center}
		\includegraphics[scale=.6]{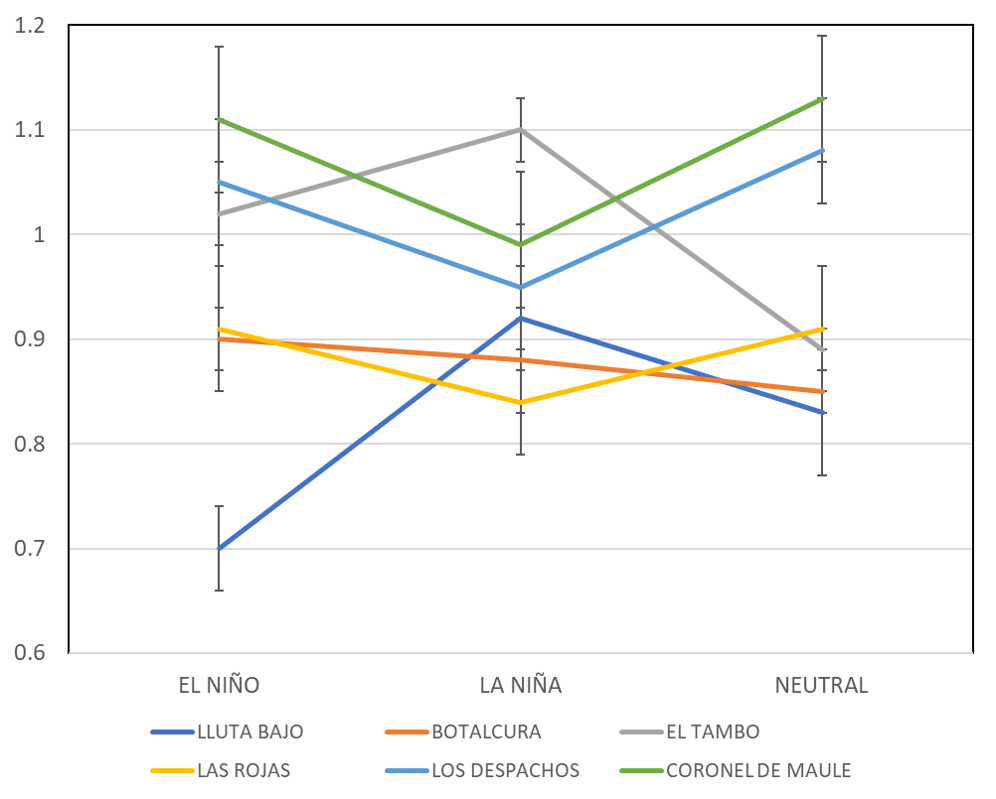}
	\end{center}
	\caption{Chilean solar radiation range for a time scale window greater than ten days. The error bar represents one standard error of the DFA coefficient.}
	\label{fig:DFA_chile_ranges}
\end{figure}

Any evaluated city in Chile have persistent or sub-diffusive behavior for macroweather. The smallest value is $0.70\pm0.04$, in Lluta Bajo. El Tambo, Los Despachos and Coronel de Maule presented sub-diffusive coefficients. Besides that, El Tambo moved from persistent in the Neutral period to sub-diffusive during the ENSO (Figure \ref{fig:DFA_chile_ranges}).\\

\begin{figure}[H]
	\begin{center}
		\includegraphics[scale=0.6]{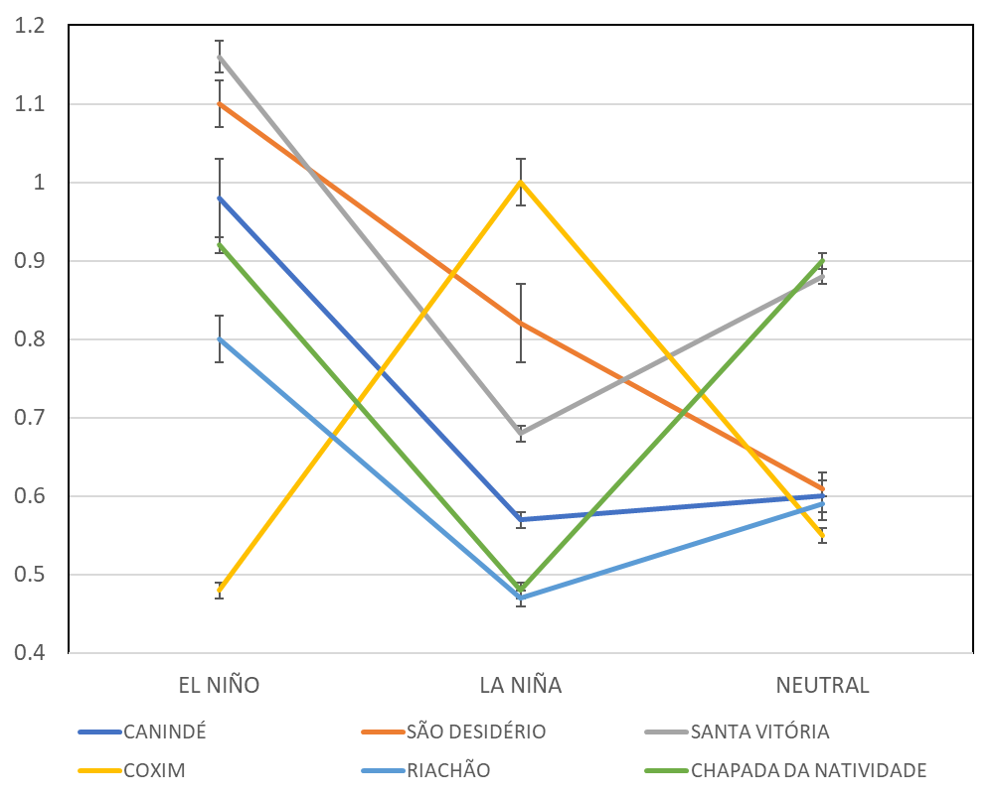}
	\end{center}
	\caption{Cities that presented a huge range variation on the $\alpha$ for solar radiation for a time scale window greater than ten days. Canind\'e, S\~ao Desid\'erio and Santa Vit\'oria where impacted by the El Ni\~no season, moving from a persistent state to a sub-diffusive one. Coxim moved from a anti-persistent state to a sub-diffusive during the La Ni\~na season. Chapada da Natividade where impacted by the El Ni\~na, and reached the anti-persistent signal $\alpha$ value. The error bar represents one standard error of the DFA coefficient.}
	\label{fig:DFA_huge_ranges}
\end{figure}

Moreover, it was noticed that some cities presented a huge range variation on the $\alpha$ for solar radiation (Figure \ref{fig:DFA_huge_ranges}). Coxim stand out in this evaluation, presenting the largest range on the $\alpha$, changing from a anti-persistent state on the El Ni\~no to a sub-diffusive one during the La Ni\~na, and become persistent in the neutral year, for a long range time scale window (Figure \ref{fig:COXIM_DFA_GRAPH}).\\

The sub-diffusive process, for scale above $10$ days, indicates a dynamic behavior of the solar irradiation. It may be compared to a transition state or transient conditions, similar to that observed in \cite{nascimento2017,nascimento2018}. This characteristic interfere the predictability in the long-term evaluation of solar radiation, as well as take decisions based on those datasets. It may impact on the Typical Meteorological Year, that is used in the calculation of CEPE.\\

\begin{figure}[H]
	\begin{center}
		\includegraphics[scale=0.6]{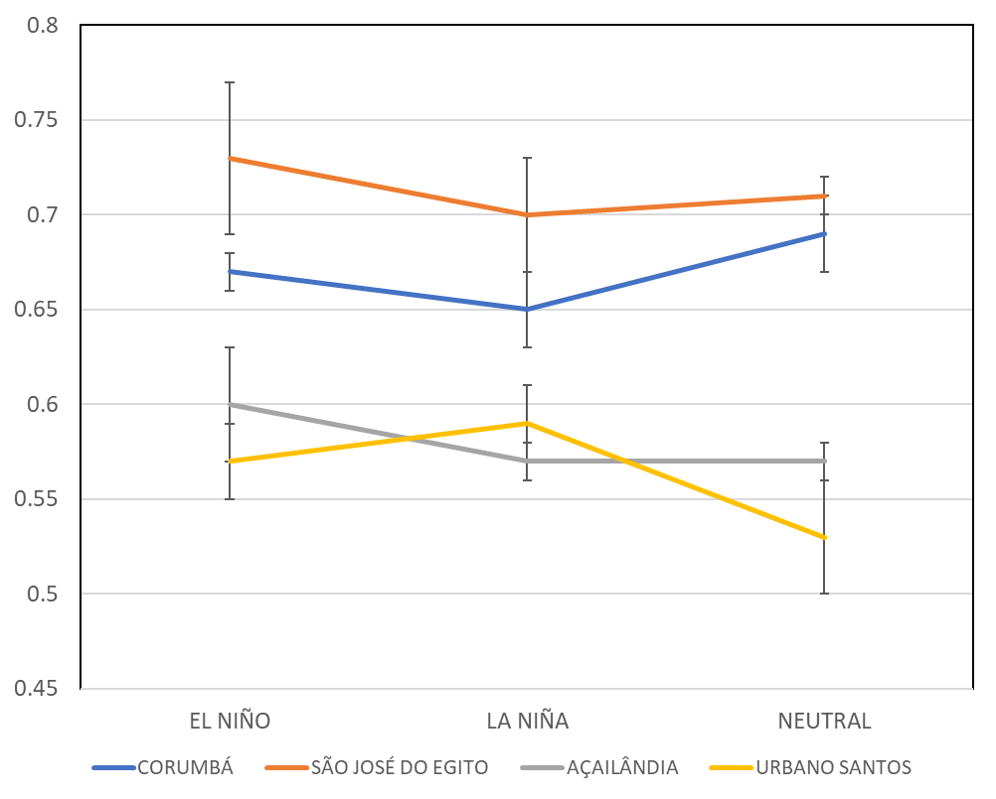}
	\end{center}
	\caption{Cities that presented small range variation on the $\alpha$ for solar radiation for a time scale window greater than ten days. Corumb\'a, S\~ao Jos\'e do Egito, A{\c c}ail\^andia e Urbano Santos have a persistent signal. The $\alpha$ range through all studied seasons did not exceed $0.05$. It means some stability of the evaluated dataset. The error bar represents one standard error of the DFA coefficient.}
	\label{fig:DFA_stable}
\end{figure}

\begin{figure}[H]
	\begin{center}
		\includegraphics[scale=0.35]{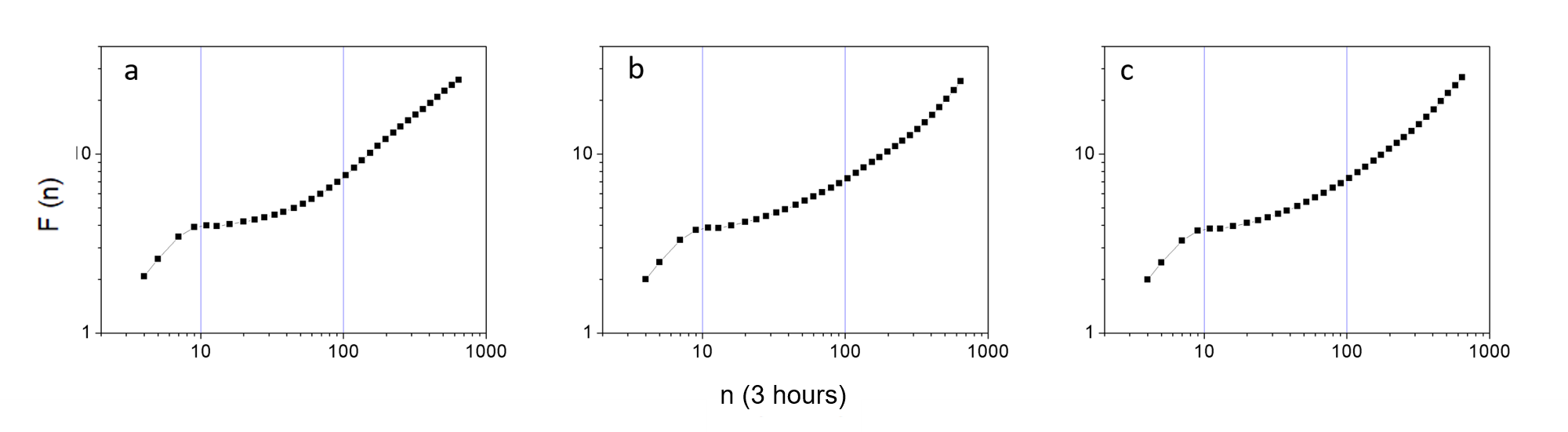}
	\end{center}
	\caption{Corumb\'a (MS) is the most stable city between the evaluated ones. It presented the smallest variation on the $\alpha$ for solar radiation during the (a) El Ni\~no, (b) La Ni\~na and (c) neutral seasons, for any time scale window.}
	\label{fig:CORUMBA_DFA_GRAPH}
\end{figure}

In contrast, this same period have some cities that presented a small range variation on the $\alpha$ (Figure \ref{fig:DFA_stable}). Corumb\'a is the most stable city between the thirty one evaluated cities. It presented the smallest range variation on the $\alpha$ for solar radiation during the (a) El Ni\~no, (b) La Ni\~na and (c) neutral seasons, for any time scale window (Figure \ref{fig:CORUMBA_DFA_GRAPH}).\\

We used a statistical equivalence test for means with paired observations to evaluate the whole series. El Niño presented a higher mean than Neutral and La Niña for a confidence level of 0.05. We cannot claim that La Niña mean is higher than Neutral. It reveals a statistically significant difference of El Niño DFA coefficients, systematically more persistent than the others.\\

In order to demonstrate the existence of spatial patterns of the alpha values, the persistence range were split in 2 steps (lower range of persistence = $0.50 > \alpha > 0.75$; higher range of persistence = $0.75 \geq \alpha > 1$). The Brazilian map were split in Above Dashed Line (ADL) and Below Dashed Line (BDL), only for visual purposes.\\

\begin{figure}[H]
	\begin{center}
		\includegraphics[scale=.9]{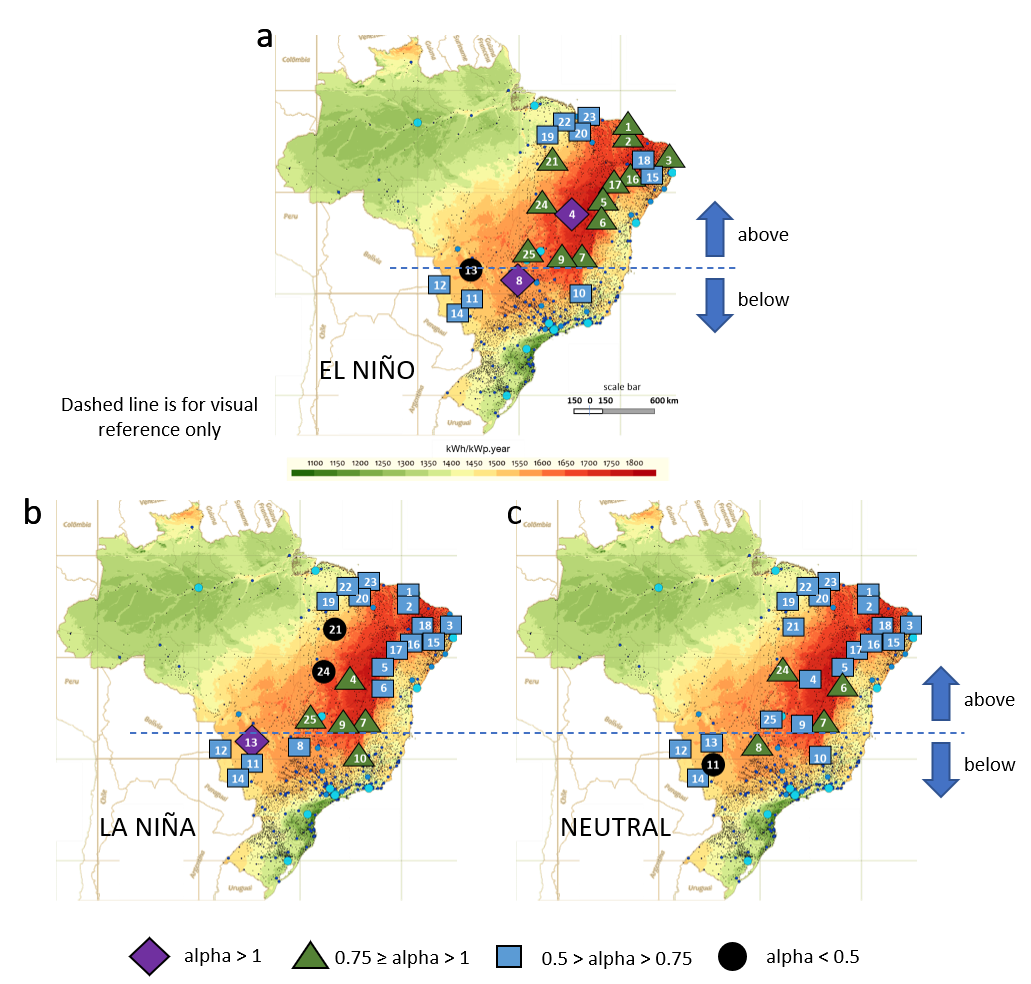}
	\end{center}
	\caption{Spatial evaluation of $\alpha$ values for El Niño (a), La Niña (b) and Neutral (c) periods in Brazil. The map were split in ADL and BDL, only for visual purposes. The most of alpha values changed from lower range to the higher range of persistence ADL during the El Niño.}
	\label{fig:Pattern_Brazil}
\end{figure}

Considering the Brazilian neutral period as baseline (Figure \ref{fig:Pattern_Brazil}c), the most of alpha values changed from lower range to the higher range of persistence ADL during the El Niño (Figure \ref{fig:Pattern_Brazil}a). The La Niña period didn`t present the same effect ADL (Figure \ref{fig:Pattern_Brazil}b). BDL, both El Niño and La Niña didn`t presented a clear pattern of changes from baseline.\\

\begin{figure}[H]
	\begin{center}
		\includegraphics[scale=1]{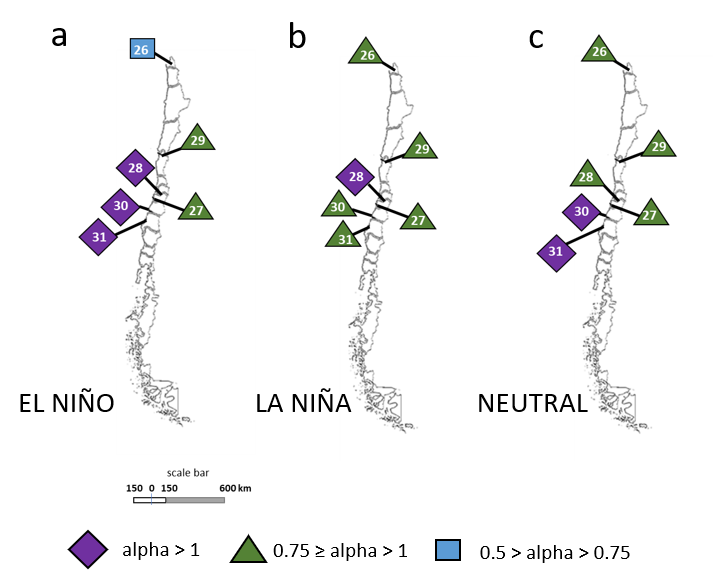}
	\end{center}
	\caption{Spatial evaluation of $\alpha$ values for El Niño (a), La Niña (b) and Neutral (c) periods in Chile. La Niña moved the alpha value from subdiffusive to persistent for the cities Los Despachos and Coronel de Maule, the southernmost analyzed cities.}
	\label{fig:Pattern_Chile}
\end{figure}

El Niño also affected the north of Chile changing the city result from higher range on neutral period \ref{fig:Pattern_Chile}c) to lower range persistence (Figure \ref{fig:Pattern_Chile}a). La Niña moved the alpha value from subdiffusive to persistent for the cities Los Despachos and Coronel de Maule, the southernmost cities among those analyzed (Figure \ref{fig:Pattern_Chile}b).\\

\section{Conclusions and perspectives}

In this paper the self-affinity of the time series for solar radiation in South America was analyzed. The neutral period is characterized mainly by the persistent behavior, determined as a desired state. But El Ni\~no and La Ni\~na showed some variation on the DFA coefficient, $\alpha$, sometimes going from persistent to anti-persistent or sub-diffusive in the same city. It means that ENSO impacts the behavior of time series of solar radiation in South America.\\

We found that El Niño is systematically more persistent than the other periods. The impact is visually homogeneous ADL on the Figure \ref{fig:Pattern_Brazil}. This spatial pattern may be compared to rainfall in Brazil, where North-Northeast of Brazil experiencing an increase in the precipitation and the South becomes dry during the La Ni\~na, while the opposite occurs in El Ni\~no. Also, the results showed the presence of crossover, a similar behavior found in other atmospheric variables in the region. La Niña impacted the Chilean cities of Los Despachos and Coronel de Maule, changing the $\alpha$ from subdiffusive to persistent.\\

The sub-diffusive process indicates a dynamic series, like a transition state or transient condition. This condition may affect the evaluation of the solar plant efficiency forecast if the measurements for CEPE were collected during ENSO. The different behaviors found, for the same location, can lead to an assessments that do not represent the long term solar energy potential generation, estimated in a period of ten years. It is a new implication for the prediction of large scale generation of solar energy. This evaluation using DFA may be complementary to the other methods already in use, in order to validate the collected data.\\

Initially, it may be also suggested to avoid the ENSO period to collect the local data for CEPE. On the other hand, future studies should focus on the use of complex methods and development of computational models to improve CEPE, trying to correlate neutral periods with ENSO modified behavior. Governments may update their energy policies for solar plant projects, based on this climatic behavior. Chile, for example, obligate new solar plant projects to be competitive in a free market \cite{grageda2016}, but a forecast based on a subdiffusive behavior may lead to a wrong decision.\\

To the best of our knowledge, it is the first time that were evaluated the self-affinity of solar radiation, in a large area of South America, to reveal the changes in the time series fluctuation, affecting the climatic behavior of the region, due ENSO. This evaluation is a start point to understand ENSO effects on the solar plant project, comparing these time series in same location. Future studies may evaluate the cross-correlation between solar radiation, temperature, humidity and wind time series, during ENSO and neutral periods, for each weather station.\\

\section{Acknowledgments}

This work received financial support from National Counsel of Technological and Scientific Development, CNPq (grant number 305291/2018-1)


\bibliography{mybibfileDFA}

\end{document}